\documentclass[reprint]{revtex4-1}

\usepackage{amssymb}
\usepackage{amsmath}
\usepackage{array}
\usepackage{dcolumn}
\usepackage{hyperref}
\usepackage{graphicx}

\usepackage{footnote}

\begin{document}
\title{Experimental Determination of the Antineutrino Spectrum\\ of the Fission Products of $^{238}$U}
\date{ \today}

\author{N. Haag}
\email[Corresponding author: ]{nils.haag@ph.tum.de}

\author{A. G\"utlein}
\author{M. Hofmann}
\author{L. Oberauer}
\author{W. Potzel}
\author{K. Schreckenbach}
\thanks{deceased}
\affiliation{Physik Department, Technische Universit\"{a}t M\"{u}nchen, 85748 Garching, Germany}
\author{F.\,M. Wagner}
\affiliation{Forschungs-Neutronenquelle Heinz Maier-Leibnitz (FRM\,II), Technische Universit\"{a}t M\"{u}nchen, 85748 Garching, Germany}

\begin{abstract}
 An experiment was performed at the scientific neutron source FRM\,II in Garching to determine the cumulative antineutrino spectrum of the fission products of $^{238}$U. This was achieved by irradiating target foils of natural uranium with a thermal and a fast neutron beam and recording the emitted $\beta$\,-\,spectra with a gamma-suppressing electron-telescope. The obtained $\beta$\,-\,spectrum of the fission products of $^{235}$U was normalized to the data of the magnetic spectrometer BILL of $^{235}$U. This method strongly reduces systematic errors in the $^{238}$U measurement. The $\beta$\,-\,spectrum of $^{238}$U was converted into the corresponding antineutrino spectrum. The final $\bar\nu_e$\,-\,spectrum is given in 250\,keV bins in the range from 2.875\,MeV to 7.625\,MeV with an energy-dependent error of 3.5\,\% at 3\,MeV, 7.6\,\% at 6\,MeV and $\gtrsim$\,14\,\% at energies $\gtrsim$\,7\,MeV (68\,\% confidence level). Furthermore, an energy-independent uncertainty of $\sim$\,3.3\,\% due to the absolute normalization is added.
Compared to the generally used summation calculations, the obtained spectrum reveals a slight spectral distortion of $\sim$\,10\,\% but returns the same value for the mean cross section per fission for the inverse beta decay.
\end{abstract}

\maketitle

\section{Introduction}
Precise predictions of the antineutrino spectra emitted by nuclear reactors are a crucial input for many current and future neutrino experiments. Apart from experiments searching for diffuse supernova neutrino events \cite{LENA} or geo-neutrinos \cite{BXGeo,KLGeo}, where reactor antineutrinos may form a substantial background to the expected signal, the knowledge of the spectrum produced by a fuel assembly is of special importance for reactor neutrino disappearance experiments. There exists a wide variety of reactor neutrino experiments aiming at the determination of neutrino (oscillation) parameters, e.g., by the current Daya Bay \cite{DayaBay2013}, RENO \cite{Reno} and Double Chooz \cite{DC2} collaborations, as well as plans for the identification of the neutrino mass hierarchy with the JUNO \cite{DB2} or Reno-50 \cite{Reno50} detectors. However, even these setups designed to compare data from near and far detectors are not fully independent of the knowledge of the emitted reactor antineutrino spectrum. Furthermore, there are questions like the possible existence of sterile neutrinos \cite{Anomaly} that cannot be studied using this comparative measurement technique. The only possibility to interpret the data of the many short baseline (up to $\sim$\,100\,m) neutrino experiments performed so far in terms of a sterile neutrino analysis is the accurate prediction of the antineutrino spectrum emitted by the particular fuel assembly.\\
Antineutrino detectors can also be used for the purposes of non-proliferation of nuclear weapons \cite{IAEA}. Monitoring the fuel composition with neutrinos from outside the reactor containment and without input from the reactor staff may give a handle to reduce the undetected manipulation and removal of nuclear fuel. In this case, the neutrino spectrum emitted by the reactor is, again, an important input to the analysis of the data.\\
In a standard pressurized water reactor (PWR), four main fuel isotopes contribute to the total power and thus to the neutrino output: $^{235}$U, $^{238}$U, $^{239}$Pu, and $^{241}$Pu. The neutron-rich fission products of these isotopes undergo beta decays emitting antineutrinos \cite{Mueller}:
\begin{eqnarray}
 S_{\bar\nu,tot}(E) = \sum_i FR_i \cdot S_{\bar\nu,i}(E)
\end{eqnarray}
with $S_{\bar\nu,tot}(E)$ being the total antineutrino spectrum emitted by the reactor core, \textit{i} representing the four main fuel isotopes, $FR_i$ being the fission rate of isotope \textit{i} and $S_{\bar\nu,i}(E)$ the total antineutrino spectrum emitted after the fission of isotope \textit{i}, including all decay spectra of the daughter isotopes.
There exist two different ways to predict the total antineutrino spectra: \\
1) The \textit{summation method} \cite{Mueller, Vogel1, Vogel} uses available databases to build up the $\beta$\,-\,spectra S$_{\beta,i}$(E) as a sum of the branch-level $\beta$\,-\,spectra of all daughter isotopes. Of course, this requires accurate knowledge of parameters like fission yields or branching ratios and of the $\beta$\,-\,spectra involved. The conversion of the final $\beta$\,-\,spectra into the antineutrino spectra $S_{\bar\nu,i}(E)$ can be performed on the branch-level with high accuracy \cite{VogelConversion, Huber}.\\
2) In the \textit{conversion method}, the four $\beta$\,-\,spectra S$_{\beta,i}$(E) of the main fuel isotopes are measured directly - without the knowledge of the underlying branch-level processes. The conversion into the $\bar\nu_e$\,-\,spectra, however, is more difficult than in the summation technique and will be described later.\\
Both methods to determine the antineutrino spectrum need the fission rates of the fuel isotopes as an input parameter, which can be calculated with reactor evolution codes like, e.g., MURE \cite{MURE} or Dragon \cite{Dragon}.\\
\noindent The accuracy of the spectra obtained by the summation method suffers from incomplete data of the $\beta$\,-\,spectra, and there is an intense discussion about the influence, for example, of the pandemonium effect \cite{Pandemonium, Fallot}, the impact of weak-magnetism corrections \cite{Huber} and of the unknown shape of many contributing $\beta$\,-\,spectra to the final calculated spectrum \cite{Hayes}. The advantage of the conversion technique is that the direct measurement of S$_{\beta,i}$(E) has no dependence on these parameters.\\
In the 1980's, measurements of the summed $\beta$\,-\,spectra S$_{\beta,i}$(E) of the fission products of three of the main fuel isotopes ($^{235}$U, $^{239}$Pu, $^{241}$Pu) were performed with the BILL magnetic spectrometer at the ILL in Grenoble \cite{BILL1,BILLU5,BILLPu}. From these spectra, the particular $\bar\nu$\,-\,spectra were derived (see also \cite{Huber}). These spectra can act as benchmark for the summation calculations, but primarily these are direct input for the conversion technique. However, as in the BILL measurement only thermal neutrons could be used, the spectrum of $^{238}$U was not measured. Consequently, until now the determination of the total $\bar\nu$\,-\,spectrum from reactors is based on the BILL data for $^{235}$U, $^{239}$Pu, and $^{241}$Pu, but has to rely on the summation calculations for $^{238}$U. This isotope contributes about 10\,\% to the total $\bar\nu_e$ output of a PWR.\\
At the scientific neutron source FRM\,II in Garching, we performed an experiment to determine - for the first time - the antineutrino spectrum of the fission products of $^{238}$U using a fast neutron beam \cite{MyThesis}. This letter describes this experiment and the resulting $\beta$\,-\,spectrum and presents a conversion into the antineutrino spectrum. \\

\section{\textbf{The $^{238}$U experiment}}

At the 20\,MW scientific neutron source FRM\,II, beam tube SR10 alternatively provides both, a fast and a thermal neutron beam \cite{FRM2}: The so-called \textit{converter facility} in front of the entrance tip of SR10 consists of two plates of highly enriched $^{235}$U with a mass m($^{235}$U)\,$\approx$\,500\,g, situated at the inner rim of the moderator tank. A fast neutron spectrum is emitted in the fission processes of this converter. In the experiment, various collimators and filters assured that no thermal neutron content was present in the fast neutron beam. Alternatively, the converter plates can be removed remotely from the moderator, resulting in a thermal neutron beam at the experiment, without the need to change the experimental setup.\\
Two identical target foils from natural uranium (99.3\,\% $^{238}$U, 0.7\,\% $^{235}$U) were irradiated, one by the thermal neutron beam and one by the fast neutron beam. In the thermal neutron beam measurement, only fission of $^{235}$U was induced, whereas with the fast neutron beam mostly fission of $^{238}$U occurred. It was thus possible to record $\beta$\,-\,spectra of the fission products of $^{235}$U and $^{238}$U with the same setup. As explained later, the $^{235}$U measurement was normalized to the spectrum of $^{235}$U obtained by the BILL experiment \cite{BILLU5} to minimize systematic uncertainties in the $^{238}$U analysis.\\
Similar to the experiment set up by Carter already in 1959 \cite{Carter}, the detector was a gamma-suppressing $\beta$\,-\,telescope consisting of two modules: 1) a spectroscopic module for full energy absorption, including a plastic scintillator and a photomultiplier, and 2) a multi-wire chamber (MWC), placed directly in front of the entrance area of the scintillator. These two modules were operated in coincidence as, due to the low density of the counting gas (CF$_4$), the MWC was not sensitive to gamma radiation also emitted by the fission and decay processes in the target foil. This suppression of gamma-induced events could be determined with $\gamma$ sources to be better than 99.5\,\%.\\
For calibration purposes, a $^{207}$Bi source and two target foils, one from polyvinylidene chloride (PVDC) and one from natural In were used. The Bi source was installed in a way that allowed to either place the source in a passive rest position (during neutron irradiation of the targets) or directly beneath the target foil. Consequently, no exchange of a target foil was necessary and any calibration with the mono-energetic conversion lines of $^{207}$Bi at $\sim$\,1\,MeV could be performed with exactly the same condition as the uranium measurements before and after this calibration run. The two calibration foils were consecutively activated by thermal neutrons, producing the beta-emitters $^{116}$In and $^{38}$Cl with endpoints at $\sim$\,3.3\,MeV and $\sim$\,4.9\,MeV, respectively, and placed at the position of the uranium target foils. The energy response was linear over the whole energy range, and an error of less than one percent on the energy calibration was reached.\\
The response function of the setup was determined by a Geant4 simulation and cross-checked at 1\,MeV by a comparison of the simulation and the measured results of the $^{207}$Bi spectrum. The resolution of the system was 8\,\%\,$\cdot\frac{1}{\sqrt{E [MeV]}}$ (FWHM).\\
To minimize effects of not having reached radioactive equilibrium, data of the first 11 hours of irradiation was not included in the analysis. Only data obtained in the following 42 hours was used.
A detailed background analysis and subtraction was executed for the fast-beam and the thermal-beam measurements separately.\\

\subsection{\textbf{Background in fast neutron beam data}}
\begin{figure}
\includegraphics[width=\columnwidth]{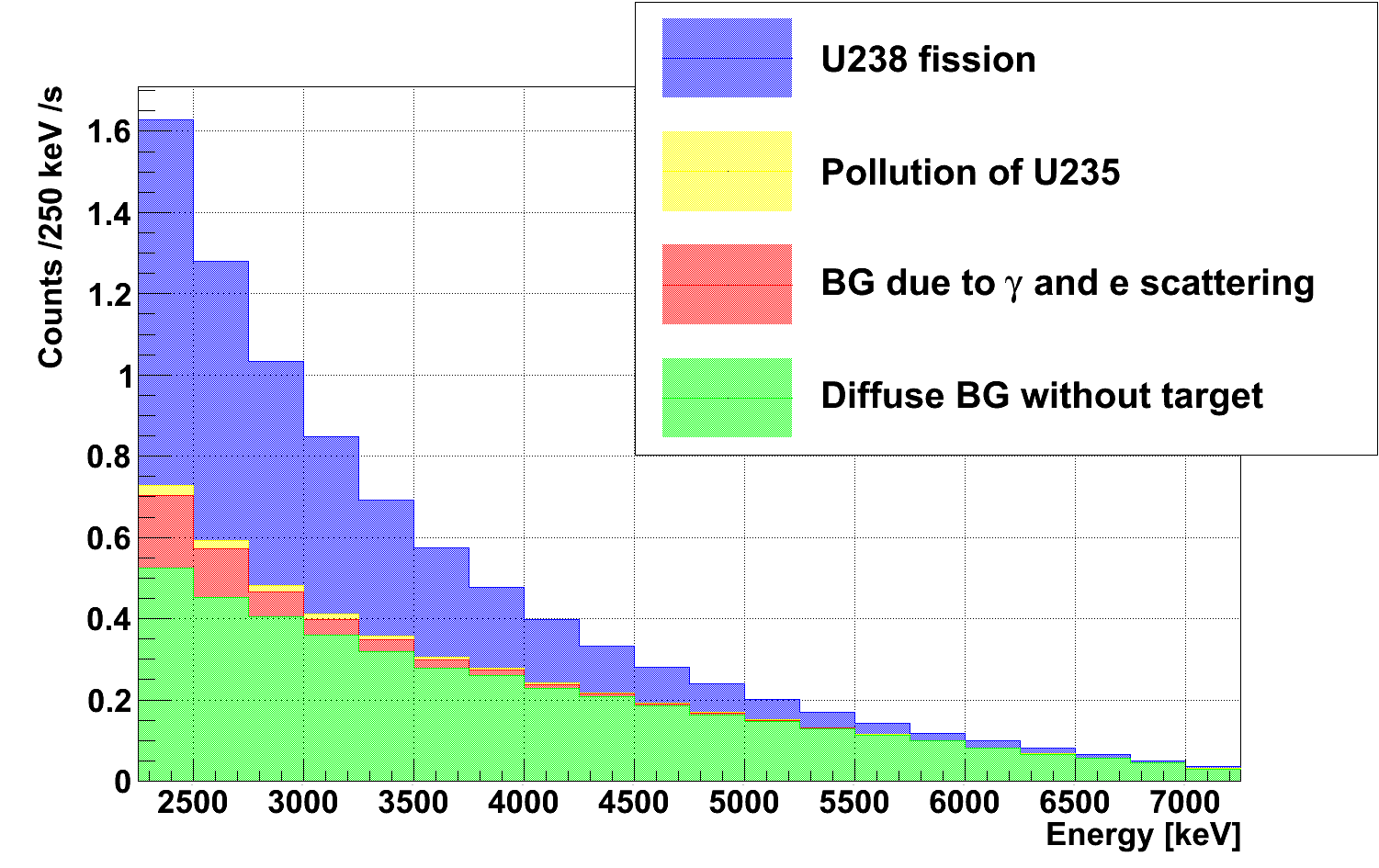} 
\caption{\textit{Stacked plot of the different contributions to the $\beta$\,-\,spectrum emitted by the uranium target foil under irradiation with fast neutrons. See text for explanation.}}
\label{Fig:BGfast}
\end{figure}

The composition of the spectrum recorded with an uranium target irradiated by the fast neutron beam was analyzed by background measurements. The latter were performed either without any target in the setup or with special dummy-targets. Figure \ref{Fig:BGfast} illustrates the results of this analysis. The main background to the uranium $\beta$\,-\,spectrum was generated by \textit{diffuse background}, which is the spectrum one obtains with the fast neutron beam on-line but without any target foil. The events in the diffuse background mainly stem from a) gammas and electrons present in the neutron beam that scatter off the collimator and the material near the detector, leading to electrons reaching the detector, b) neutrons captured in the vicinity of the detector producing instable isotopes and high-energy gamma radiation and c) cosmic muons. This diffuse target-independent background had an intensity similar to the $\beta$\,-\,spectrum from the fission products of $^{238}$U, however, it could be determined with high accuracy by a ten hour measurement. Nevertheless, this background caused the statistical error to exceed $\sim$\,10\,\% at energies greater than $\sim$\,6.5\,MeV.\\
In addition, scattered electrons and gammas converted into electrons at the target foils cause a target-dependent background. To determine this background present in the uranium data, measurements with dummy targets of lead and nickel were performed. With the help of Geant4 simulations, the background spectra recorded with the dummy targets could be used to calculate the background spectra present in the uranium data.\\
Finally, as the uranium targets consisted of natural uranium, fission of $^{235}$U contributed to the fast neutron beam spectrum. Due to the knowledge of the neutron beam spectrum, of the fission cross sections and the measurement of the $^{235}$U in the thermal beam, this small contribution (2.6\,\% in total) could be corrected for.\\

\subsection{\textbf{Background in thermal neutron beam data}}
As the converter plate was removed, the gamma and electron content present in the neutron beam - and thus the diffuse background - was significantly reduced. The background in the thermal beam was dominated by captures of thermal neutrons by the material surrounding the detector module, leading to electrons and high-energy gammas that were partly converted into electrons. This contribution, as well as the remaining background induced by gammas from the beam, could again be determined with the help of dummy target measurements (Pb and Ni). The signal to background ratio was 6.5 at 4\,MeV and better than 1 up to energies of 6.5\,MeV.\\

\subsection{\textbf{Normalisation to the BILL measurement}}
The advantage of the experiment described herein was the possibility to normalize the data to the very accurate BILL measurement of the spectrum of the fission products of $^{235}$U. This is of paramount importance as it greatly suppresses systematic uncertainties connected to, e.g.,  the unknown detector efficiency, unexpected energy-dependence of the detector response function, and the barely known neutron beam profile and intensity.\\
A normalization function NF is defined bin-wise as the ratio of the $\beta$\,-\,spectra of $^{235}$U measured in the present experiment (U235) and the BILL experiment (BILL). NF is then applied to the $^{238}$U spectrum:
\begin{equation}
 NF := \frac{U235}{BILL} = F_{\gamma} \frac{U238}{U238_{final}}
\end{equation}
\begin{equation}
\Rightarrow U238_{final} = F_{\gamma} \cdot U238\cdot \frac{BILL}{U235}
\label{Eq:NF}
\end{equation}
Herein, \textit{U}238 is the $\beta$\,-\,spectrum of the fission products of $^{238}$U, as measured in the present experiment (after background subtraction), and \textit{U}238$_{final}$ is the $^{238}$U spectrum quoted as final result. As this normalization did not take into account the different number of fissions in the two target foils during irradiation, the factor $F_{\gamma}$ had to be introduced. This factor was determined by $\gamma$\,-\,spectrometry of the irradiated foils after the end of the measurement campaign ($F_{\gamma}$ = 44.4\,$\pm$\,0.3) by measuring the peak areas of selected gamma lines emitted by the fission products with a high-resolution germanium spectrometer \cite{HofmannPhD}.\\
This normalization fully correlates the $^{238}$U spectrum obtained in this present experiment with the BILL spectrum of $^{235}$U (see eq. \ref{Eq:NF}). 

\begin{table}[]
 \centering
	\begin{tabular}{|c|c|c|c|c|}
	\hline
	 E [keV]& N$_{\beta}$ $\left[ \frac{betas}{fiss. \cdot MeV}\right]$& $\epsilon$ [\%]& $\epsilon_{exp,norm}$ [\%]&$\epsilon_{BILL}$ [\%]\\
	\hline
	\hline
	2250\,-\,2500	& 1.032		&3.2 	& 2.1 & 1.7\\
	2500\,-\,2750	& 8.302 $\cdot$ 10$^{-1}$& 3.0	& 2.1& 1.7\\
	2750\,-\,3000	& 6.922 $\cdot$ 10$^{-1}$& 2.4	& 2.1& 1.7\\
	3000\,-\,3250	& 5.698 $\cdot$ 10$^{-1}$& 2.3	& 2.1& 1.7\\
	3250\,-\,3500	& 4.533 $\cdot$ 10$^{-1}$& 2.4	& 2.1& 1.7\\
	3500\,-\,3750	& 3.740 $\cdot$ 10$^{-1}$& 2.4	& 2.1& 1.7\\
	3750\,-\,4000	& 2.807 $\cdot$ 10$^{-1}$& 2.7	& 2.1& 1.7\\
	4000\,-\,4250	& 2.279 $\cdot$ 10$^{-1}$& 2.9	& 2.1& 1.7\\
	4250\,-\,4500	& 1.725 $\cdot$ 10$^{-1}$& 3.5	& 2.1& 1.8\\
	4500\,-\,4750	& 1.343 $\cdot$ 10$^{-1}$& 3.9	& 2.1& 1.8\\
	4750\,-\,5000	& 1.084 $\cdot$ 10$^{-1}$& 4.5	& 2.1& 1.8\\
	5000\,-\,5250	& 7.891 $\cdot$ 10$^{-2}$& 5.5	& 2.1& 1.8\\
	5250\,-\,5500	& 5.831 $\cdot$ 10$^{-2}$& 6.8	& 2.1& 1.8\\
	5500\,-\,5750	& 4.137 $\cdot$ 10$^{-2}$& 9.7	& 2.1& 1.8\\
	5750\,-\,6000	& 2.909 $\cdot$ 10$^{-2}$& 11.7	& 2.1& 1.8\\
	6000\,-\,6250	& 2.765 $\cdot$ 10$^{-2}$& 11.1	& 2.1& 1.8\\
	6250\,-\,6500	& 2.248 $\cdot$ 10$^{-2}$& 12.7	& 2.1& 1.8\\
	6500\,-\,6750	& 1.296 $\cdot$ 10$^{-2}$& 18.9	& 2.1& 1.9\\
	6750\,-\,7000	& 7.078 $\cdot$ 10$^{-3}$& 28.1	& 2.1& 1.9\\
	\hline
	\hline	
	\end{tabular}
		\caption{\textit{The final result of the measurement of the $\beta$\,-\,spectrum of the fission products of $^{\text{238}}$U. N$_{\beta}$ is the $\beta$\,-\,spectrum given in units of betas per fission and MeV. The relative combined statistical and systematic error $\epsilon$ is given in column 3. Column 4 gives the error $\epsilon_{exp,norm}$ which is due to uncertainties in the absolute scale normalization of the present experiment. The last column shows the error $\epsilon_{BILL}$ on the absolute rate in the BILL measurement. All errors given at 68\,\% confidence level (1\,$\sigma$).}}
		\label{Tab:U238BetaData}
\end{table}
The final $\beta$\,-\,spectrum of $^{238}$U is given in table \ref{Tab:U238BetaData}.
The error $\epsilon$ quoted in the third column of table \ref{Tab:U238BetaData} is the energy-dependent, bin-to-bin uncorrelated, combined statistical and systematic error of the measurement. In addition to $\epsilon$, there is an almost energy-independent uncertainty of the absolute normalization $\epsilon_{norm}$\,$\approx\,2.8$\,\%, calculated as the quadratic sum of $\epsilon_{exp,norm}$ and $\epsilon_{BILL}$. $\epsilon_{exp,norm}$ is due to inaccuracies of the germanium spectrometry and different dead-time corrections for the two uranium measurements and $\epsilon_{BILL}$ is the error of absolute scale in the BILL experiment. The latter adds a slight energy-dependence to the absolute normalization error and is listed separately to be able to disentangle the different contributions when using another normalization than the BILL spectrum would be desired. A lower threshold of 2250\,keV had to be set to avoid an additional, in the present experiment indeterminable background of the 2.2\,MeV gammas from the capture of neutrons by the scintillation detector itself \cite{MyThesis}.

\subsection{\textbf{Conversion to antineutrino spectrum}}
As last step, the obtained $\beta$\,-\,spectrum was converted into an $\bar\nu_e$\,-\,spectrum. A standard conversion procedure applied to the BILL spectra is based on introducing hypothetical beta branches which are used to fit the experimental $\beta$\,-\,spectra and subsequently are converted on the branch level into the corresponding neutrino spectra. From these, the total neutrino spectrum is built up. Due to the low statistics in the high-energy regime, this technique could not be used for the $^{238}$U measurement presented herein. Instead, an empirical method, already proposed in \cite{BILLU5}, was chosen: Being the sum spectra of many decaying isotopes, the cumulative $\beta$\,- and $\bar\nu_e$\,-\,spectra emitted are very similar. The main differences in the spectra can be corrected for by shifting the electron spectrum by 511\,keV (the mass of the electron) and, in addition, a smaller amount of 50\,keV to account for an average Coulomb-attraction of a nucleus and the charged electron \cite{Vogel}. All further corrections - which are of the order of 5\,\% \cite{MyThesis} - can be described by a factor k(E):
\begin{eqnarray}
 N_{\bar\nu}(E) = N_{\beta}(E - 511 keV - 50 keV) \cdot k(E)
\end{eqnarray}

This factor k(E) could be extracted from the BILL measurements and the predictions from the summation method. Note that even though the summation predictions result in relatively high errors of the final spectra - partly due to unknown $\beta$\,-\,spectra - the factor k(E) extracted from these is more reliable as it is only sensitive to errors in the branch-level conversion of the $\beta$\,- to the $\bar\nu_e$\,-\,spectra.\\
Table \ref{Tab:U238NuData} gives the final antineutrino spectrum of the fission products of $^{238}$U as determined by the present experiment.
 
\begin{table}[]
 \centering
	\begin{tabular}{|c|c|c|c|}
	\hline
	 E [keV]& N$_{\bar\nu}$ $\left[ \frac{\bar\nu}{fission \cdot MeV}\right]$& $\epsilon$ [\%]& $\epsilon_{norm}$ [\%]\\
	\hline
	\hline
	3000	& 9.586 $\cdot$ 10$^{-1}$&3.5 	& 3.3\\
	3250	& 7.952 $\cdot$ 10$^{-1}$& 3.1	& 3.3\\
	3500	& 6.603 $\cdot$ 10$^{-1}$& 2.6	& 3.3\\
	3750	& 5.406 $\cdot$ 10$^{-1}$& 2.6	& 3.3\\
	4000	& 4.433 $\cdot$ 10$^{-1}$& 2.6	& 3.3\\
	4250	& 3.498 $\cdot$ 10$^{-1}$& 2.8	& 3.3\\
	4500	& 2.787 $\cdot$ 10$^{-1}$& 2.9	& 3.3\\
	4750	& 2.171 $\cdot$ 10$^{-1}$& 3.3	& 3.3\\
	5000	& 1.700 $\cdot$ 10$^{-1}$& 3.7	& 3.4\\
	5250	& 1.341 $\cdot$ 10$^{-1}$& 4.1	& 3.4\\
	5500	& 1.032 $\cdot$ 10$^{-1}$& 5.0	& 3.4\\
	5750	& 7.737 $\cdot$ 10$^{-2}$& 5.9	& 3.4\\
	6000	& 5.618 $\cdot$ 10$^{-2}$& 7.6	& 3.4\\
	6250	& 3.973 $\cdot$ 10$^{-2}$& 10.6	& 3.4\\
	6500	& 3.048 $\cdot$ 10$^{-2}$& 12.6	& 3.4\\
	6750	& 2.805 $\cdot$ 10$^{-2}$& 11.7	& 3.4\\
	7000	& 2.093 $\cdot$ 10$^{-2}$& 14.1	& 3.4\\
	7250	& 1.139 $\cdot$ 10$^{-2}$& 21.9	& 3.4\\
	7500	& 7.132 $\cdot$ 10$^{-3}$& 30.0	& 3.4\\
	\hline
	\hline	
	\end{tabular}
\caption{\textit{The $\bar \nu_e$\,-\,spectrum of the fission products of $^{\text{238}}U$. The energies E represent the center of the 250\,keV wide bins. $\epsilon$ is the combined inaccuracy of all error sources (statistical + systematic), except for the global absolute normalization uncertainty $\epsilon_{norm}$ which is quoted in the last column and includes the uncertainty of the BILL measurement. All errors given at 68\,\% confidence level (1\,$\sigma$).}}
		\label{Tab:U238NuData}
\end{table}

\subsection{\textbf{Discussion}}
\begin{figure}
\includegraphics[width=\columnwidth]{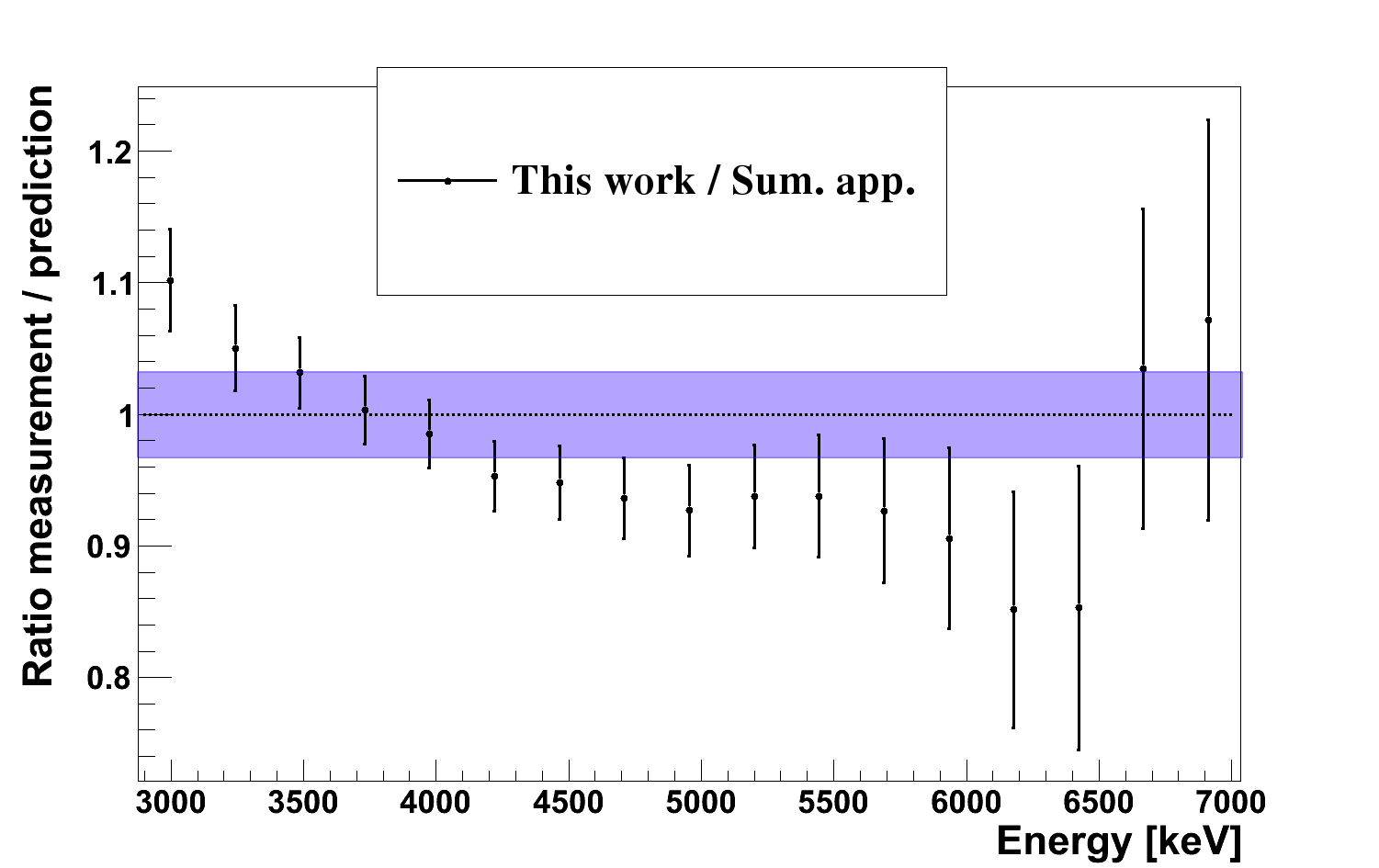} 
\caption{\textit{The ratio of the measured $\bar\nu_e$\,-\,spectrum of $^{\text{238}}$U and the result of the summation approach by \cite{Mueller}. The shaded band represents the error of 3.3\,\% of the absolute normalization and the error bars illustrate the combined systematic and statistical error of the experiment only. No error was included from the summation calculations. One can see a slight spectral distortion of the order of 10\,\%.}}
\label{Fig:Comp}
\end{figure}

Comparing the experimental $\bar\nu_e$\,-\,spectrum of the fission products of $^{238}$U with the results from the summation approach reveals a slight spectral distortion in the currently assumed shape. Figure \ref{Fig:Comp} plots the ratio of the experimental spectrum and the results of the summation method in \cite{Mueller}. This was chosen, as it is the one currently in use in the reactor neutrino disappearance experiments.\\

\noindent Furthermore, the reduced mean cross section per fission $<\sigma_f>$ for the inverse beta decay can be calculated:
\begin{eqnarray}
 <\sigma_{f,U238}>_{red} &=& \int^{7.625\,MeV}_{2.875\,MeV} S_{U238}(E)\cdot\sigma_{IBD} dE \nonumber \\
= 8.51\cdot 10^{-43} &\pm& 9.07 \cdot 10^{-45} (stat. + sys.)  \nonumber \\
&\pm& 2.80 \cdot 10^{-44} (norm.) \frac{cm^2}{fiss.}
\end{eqnarray}
with S$_{U238}$(E) being the $\bar\nu_e$\,-\,spectrum of the fission products of $^{238}$U and $\sigma_{IBD}$ the cross section for the inverse beta decay. The label \textit{red} indicates that the mean cross section could only be calculated for the reduced energy range covered by the experiment. The ratio between the reduced mean cross section determined by the experiment (Exp.) and the summation method (Sum.) from \cite{Mueller} is:

	\begin{eqnarray}
	\frac{<\sigma_{f,U238}>_{red}(Exp.)}{<\sigma_{f,U238}>_{red}(Sum.)} =	0.97 &\pm& 0.08 (sys + stat.) \nonumber \\
	&\pm& 0.03 (norm.) 
	\end{eqnarray}

The error of 0.08 (sys. + stat.) comprises the uncertainties of experiment and summation approach and is highly dominated by the error of the theoretical calculations. The uncertainty of 0.03 represents the error of the absolute normalization of the experiment. Thus, the ratio is compatible with 1 and the experiment confirms the currently assumed value for the mean cross section per fission. 

\section{\textbf{Conclusion}}

For the first time, the $\bar\nu_e$\,-\,spectrum of the fission products of $^{238}$U could be determined experimentally. The experiment covers an energy range from 2.875\,MeV to 7.625\,MeV and has a relative energy-dependent error of 3.5\,\% at 3\,MeV, 7.6\,\% at 6\,MeV and $\gtrsim$14\,\% at energies $\gtrsim$7\,MeV (68\,\% confidence level). The uncertainty of the absolute scale of $\sim$\,3.3\,\% is almost energy-independent. The results reveal spectral distortions in the currently used spectrum achieved with summation calculations of the order of 10\,\%. The value of the mean cross section per fission in the energy range covered by the experiment (2.875\,-\,7.625\,MeV) matches the one determined by the summation approach. With this spectrum, it is now possible to determine reactor antineutrino spectra without use of theoretical spectra of the contributing fission isotopes.

\section{Acknowledgments}

In commemoration of our colleague and friend Klaus Schreckenbach we would like to express our heart-felt sympathy to his family. We want to thank the staff of the FRM\,II and especially the ANTARES-team. Thanks also to M. Fallot, D. Lhuillier and their groups for the kind and fruitful discussions. This work was funded by the Cluster of Excellence 'Origin and Structure of the Universe' and the DFG (research grant no. GO1729).


\begin{thebibliography}{xxxxxxx}
 	\addcontentsline{toc}{chapter}{Bibliography}

	\bibitem[1]{LENA} \textsc{M.\,Wurm et al.}, Astropart. Phys. 35, 685 (2012)

	\bibitem[2]{BXGeo} \textsc{G.\,Bellini et al.}, Phys. Lett. B\,722, 295 (2013)

	\bibitem[3]{KLGeo} \textsc{A.\,Gando et al.}, Nat. Geosc. 4, 647 (2011)

	\bibitem[4]{DayaBay2013} \textsc{F.\,P.\,An et al.}, Chin. Phys. C\,37, no. 1, 011001 (2013)

	\bibitem[5]{Reno} \textsc{J.\,K.\,Ahn et al.}, Phys. Rev. Lett. 108, 191802 (2012)

	\bibitem[6]{DC2} \textsc{Y.\,Abe et al.}, Phys. Rev. D\,108, 052008 (2012)

	\bibitem[7]{DB2} \textsc{Y.\,-\,F.\,Li et al.}, Phys. Rev. D 88, 013008 (2013)

	\bibitem[8]{Reno50} \textsc{S.\,-\,H. Seo}, Talk at international workshop on Reno-50, Seoul, (June 2013)

	\bibitem[9]{Anomaly} \textsc{G.\,Mention et al.}, Phys. Rev. D\,83, 073006 (2011)

	\bibitem[10]{IAEA} \textsc{H.\,Furuta et al.}, IAEA-CN-184/63 (2010)

	\bibitem[11]{Mueller} \textsc{Th.\,A.\,Mueller et al.}, Phys. Rev. C\,83, 054615 (2011)

	\bibitem[12]{Vogel1} \textsc{P.\,Vogel et al.}, Phys. Ref. C\,19, 2259 (1979)

	\bibitem[13]{Vogel} \textsc{P.\,Vogel et al.}, Phys. Ref. C\,24, 1543 (1981)

	\bibitem[14]{VogelConversion} \textsc{P.\,Vogel}, Phys. Ref. C\,76, 025504 (2007)

	\bibitem[15]{Huber} \textsc{P.\,Huber}, Phys. Rev. C\,84, 024617 (2011), Errata: Phys. Rev. C\,85, 029901(E) (2012)

	\bibitem[16]{MURE} \textit{http://lpsc.in2p3.fr/MURE/pdf/UserGuide.pdf} (2012)

	\bibitem[17]{Dragon} \textsc{G.\,Marleau et al.}, Report IGE-157, Institut de g\'{e}nie nucl\'{e}aire, \'{E}cole Polytechnique de Montr\'{e}al (1994)

	\bibitem[18]{Pandemonium} \textsc{J.\,C.\,Hardy et al.}, Phys. Lett. B\,71, 307 (1977)

	\bibitem[19]{Fallot} \textsc{M.\,Fallot et al.}, Phys. Rev. Lett. 109, 202504 (2012)

	\bibitem[20]{Hayes} \textsc{A.\,C. Hayes et al.}, arXiv:1309.4146 [nucl-th] (2013)

	\bibitem[21]{BILL1} \textsc{K.\,Schreckenbach et al.}, Phys. Lett. 99\,B, 251 (1981)

	\bibitem[22]{BILLU5} \textsc{K.\,Schreckenbach et al.}, Phys. Lett. 160\,B, 325 (1985)

	\bibitem[23]{BILLPu} \textsc{A.\,A.\,Hahn et al.}, Phys. Lett. B\,218, 365 (1989)

	\bibitem[24]{MyThesis} \textsc{N.\,Haag}: \textit{''Experimental Determination of the Antineutrino Spectrum of the Fission Products of $^{\text{238}}$U''}, PhD thesis, Technische Universit\"at M\"unchen (2013), available at http://nbn-resolving.de/urn/resolver.pl?urn:nbn:de:bvb:91-diss-20131017-1171187-0-3

	\bibitem[25]{FRM2} \textsc{H. Breitkreutz et al.}, Nucl. Instr. Meth.  A593, 466 (2008)

	\bibitem[26]{Carter} \textsc{R.\,E.\,Carter et al.}, Phys. Rev. 113, 280 (1959)

	\bibitem[27]{HofmannPhD} \textsc{M. Hofmann}: \textit{''Liquid Scintillators and Liquefied Rare Gases for Particle Detectors''}, PhD thesis, Technische Universit\"at M\"unchen (2012), available at http://nbn-resolving.de/urn/resolver.pl?urn:nbn:de:bvb:91-diss-20121203-1115726-1-6 

\end{thebibliography}
\end{document}